\documentclass[apjl]{emulateapj}

\begin{document}


\title{Deep optical images of Malin 1 reveal new
  features}
\footnote{This paper includes data gathered with the 6.5
    meter Magellan Telescopes located at Las Campanas Observatory,
    Chile.}
%\title{The spiral arms of Malin 1 revealed through deep optical
%  imaging} 


\author{Gaspar Galaz\altaffilmark{1}, Carlos
  Milovic\altaffilmark{2,3,4}, 
  Vincent Suc\altaffilmark{1,5}, Luis Busta\altaffilmark{6}, Guadalupe
  Lizana\altaffilmark{1}, Leopoldo Infante\altaffilmark{1,7}, Santiago
  Royo\altaffilmark{5}}


\altaffiltext{1}{Instituto de Astrof\'{\i}sica, Pontificia Universidad
  Cat\'olica de Chile.} 
\altaffiltext{2}{Departamento de Ingenier\'{\i}a El\'ectrica, Pontificia
  Universidad Cat\'olica de Chile.} 
\altaffiltext{3}{Centro de Im\'agenes Biom\'edicas, Pontificia Universidad
  Cat\'olica de Chile.} 
\altaffiltext{4}{PixInsight Development Team, Pleiades Astrophoto
  S.L., Spain.} 
\altaffiltext{5}{Centre for the Development of Sensors, Instruments
  and Systems, UPC-BarcelonaTech.}
\altaffiltext{6}{Institute for Astro and Particle Physics, Universitat
  Innsbruck.}
\altaffiltext{7}{Centro de Astro-Ingenier\'{\i}a, Pontificia
  Universidad Cat\'olica de Chile.}


\begin{abstract}
We present Megacam deep optical images ($g$ and $r$) of Malin 1 obtained
with the 6.5m Magellan/Clay telescope, detecting structures down to $\sim 28$
$B$ mag arcsec$^{-2}$. In order to
enhance galaxy features buried in the noise, we use a
noise reduction filter based on the total generalized 
variation regularizator. This method allows us to
detect and resolve very faint morphological features, including
spiral arms, with a high visual contrast. For the first time, we can
appreciate an optical image of Malin 1 and its morphology
in full view. The images provide unprecedented detail,
compared to those obtained in the past with photographic plates and
CCD, including HST imaging. We detect two peculiar features in the
disk/spiral arms. The analysis suggests that the first one is possibly
a background galaxy, and the second is an apparent stream without a clear
nature, but could be related to the claimed past interaction between
Malin 1 and the galaxy SDSSJ123708.91 + 142253.2. Malin 1 exhibits
features suggesting the presence of stellar associations, and clumps of
molecular gas, not seen before with such a clarity. Using these
images, we obtain a diameter for Malin 1 of 160 kpc, $\sim
50$ kpc larger than previous estimates. A simple
analysis shows that the observed spiral arms reach very low luminosity
and mass surface densities, to levels much lower  
than the corresponding values for the Milky Way.
\end{abstract}


\section{Introduction}

One of the main factors precluding the observability of the
extragalactic universe is the limitation in surface brightness (SB) by different
surveys. In the seventies, different authors have shown that the
universe is populated by low surface brightness galaxies (LSBGs)
having much lower SB than the dark night sky
\citep{disney1976, bothun1987}. \citet{dalcanton1997} showed using
data from CCD drift scans from the eighties, that one can expect about
4 galaxies/deg$^2$ between the range 23-25 mag arcsec$^{-2}$. In the nineties and
already entering in the new
millennium, there were no surveys with significantly fainter surface
brightness limits. Some progress on the statistical significance of the
population of LSBGs has been made in recent years because of the 
higher volumes sampled, thanks to massive surveys like the SDSS and others, but
still barely
reaching $\sim 23.0$ mag arcsec$^{-2}$. On the other hand, and quite
unexpectedly, \citet{vandokkum2015, koda2015} reported on the
discovery of a dozen of Milky Way (MW) sized, passively evolving, ultra diffuse galaxies
(UDG) in the Coma cluster. This population is very likely dark-matter
dominated and thus represents a challenge to the current theories of galaxy
formation. Without the presence of large fractions of dark matter in
LSBGs, it is very difficult to prevent the rapid destruction of low density
galaxies within a massive cluster like Coma \citep{toomre1972,
  moore1999, mcgee2010}.    

Thus, there are still many unsolved issues concerning the nature of
LSBGs. One of these issues is the nature of the class of the so-called
giant LSBGs: large format spiral galaxies with a similar or
larger size than the MW and very low SB. Big spirals exhibiting similar
morphologies of the grand design spiral galaxies like M31 and the MW,
but with much lower stellar density, as described in
\citet{sprayberry1995, impey1996, 
  galaz2002, galaz2011}, and references therein. The
best and more extreme example in this category
is Malin 1, a disk/spiral active galaxy \citep{impey1989, barth2007} with a disk
SB of $\geq 24$ $B$ mag arcsec$^{-2}$, an uncertain inclination
(the HI data indicate that this galaxy has an inclination of about 50
deg), and a barred inner disk \citep{barth2007}. In addition to this faint 
SB, what makes Malin 1 exceptional is its size: about 110 kpc
diameter in HI and presumably a similar or larger diameter in 
the optical, making this galaxy the largest one in the universe
detected so far ($\sim 6.5$ times the MW diameter). Malin 1 was
serendipitously discovered by  
\citet{bothun1987} in photographic plates. It is close to NGC 7145 in the
Virgo cluster, and near a bright star. This last feature prevents to
obtain deep optical imaging easily without contaminating the 
FOV with scattered light, imposing additional 
difficulties for observing its morphology. Malin 1 has been studied so
far in HI \citep{pickering1997, lelli2010}, and in the optical
\citep{barth2007,  moore2006}. It is from these studies that its huge
size was measured. Also, the HI data allow to estimate a HI mass of
$\sim 10^{10}$ M$_\odot$, and a dynamical
mass of $\sim$10$^{12}$ M$_\odot$, which makes Malin 1 to be an enormous dark
matter reservoir \citep{seigar2008}. The galaxy has been 
also an interesting target for Spitzer \citep{rahman2007} and for sub-mm
observations \citep{das2006}, both for searching warm and cold dust, as well as
to detect emission from molecular clouds. These observational efforts
failed subsequently, which means that Malin 1 has a very low gas
density and/or low dust content. These observational facts place
Malin 1 at the bottom end of the Krumholz diagram
\citep{krumholz2005}, which relates the stellar formation efficiency and
the gas surface density. In this diagram Malin 1 is
certainly one of the galaxies having the lowest stellar formation
efficiency with one of the lowest molecular gas density.  

In spite of the wealth of information for Malin 1,
including HST imaging
\citep{oneil2000} and ground-based observations \citep{moore2006,
  reshetnikov2010}, optical images usually lack of good quality and are not
deep enough to detect features fainter than 24 $B$ mag
arcsec$^{-2}$. For
example, HST observations of Malin 1 \citep{oneil2000} are about 15 minutes
exposure, reaching data basically for the bulge of the
galaxy, but extremely shallow images for other structures, including
the disk and the presumed spiral arms. \citet{moore2006} built deep
imaging of Malin 1 co-adding $R$ images from 63 UKST photographic films
digitally scanned by the superCOSMOS machine, covering the astonishing area of 36
deg$^2$. \citet{reshetnikov2010} obtained 
spectroscopic observations for the central part of Malin 1 to prove
that Malin 1B, a galaxy located 14 kpc from the Malin 
1 center, is interacting with Malin 1. But they do not provide
optical imaging. Also relevant for this discussion is the work by
\citet{barth2007}, which claim that Malin 1 is a normal galaxy (with a
normal disk), embedded into a larger, more diffuse very low SB
disk. However, in his analysis, Barth uses the same \citet{oneil2000} HST
images, which are relatively shallow ones. In summary,
no deep, high quality CCD images are available for Malin 1.

In this letter we present deep optical images taken with
Megacam at the 6.5m Magellan/Clay telescope. We exposed for about 4.5 hours
in $r$ and $g$, reaching $\sim 28$ $B$ mag arcsec$^{-2}$, which allows
to observe for the first time features with unprecedented
detail. We perform a careful reduction and we use a novel image
processing in order to enhance data practically buried into the
noise. 

The paper is organized as follow. In \S 2 we present the observations
and key steps in the data reduction, in \S 3 the main
results and discussion. We conclude in \S 4. Through
this paper we use H$_0$ = 71 km s$^{-1}$ Mpc$^{-1}$; $\Omega_M = 0.27$;
$\Omega_\Lambda = 0.73$. 

\section{Observations and data reduction}

Observations were performed at the 6.5m Magellan/Clay telescope using
Megacam with filters $r$ and $g$, on the night of 25 April 2014. The night
was dark, moonless, a necessary restriction for low SB targets. The
night was also photometric with a median seeing of 
0.8 arcsec. Exposure
times were 4.8 hrs in $r$ and 4.2 hrs in $g$ (individual images of
10 minutes). After cosmic rays rejection, individual images were
flat-field and illumination corrected, then stacked, and properly
combined (see left panels of Figure \ref{fig1}). The flat-field and
illumination corrections are 
critical steps. Pixel to pixel variations along the images, as well as
illumination inhomogeneities both can introduce mid- and large scale
image artifacts that can be erroneously attributed to physical features of the
target. In addition to the low SB of Malin 1, the field where the 
galaxy is located makes difficult to obtain deep
images. First, a bright star of 8.5 mag (BD+15 2483) is
located 4.7 arcmin SW of Malin 1. Second, the galaxy NGC4571 is
located 6.6 south of Malin 1. These two 
sources generate scattered light and sky brightness 
which contaminate the already faint structure of Malin 1. 
To combine different frames, synthetic background models were extracted.
These models allow to remove large scale components preventing
a proper match between frames. Frame matching, image registration,
image combination, and all further post-processing steps were done using 
PixInsight, co-developed by one of the authors of this letter
\citep{conejero2015}. The average sky SB was 20.2
$r$ mag arcsec$^{-2}$, and 21.8 $g$ mag arcsec$^{-2}$,
equivalent to $\sim 22.5$ $B$ mag arcsec$^{-2}$, using the filter
transformations from \citet{jordi2006}.    

\subsection{Noise treatment and Multiscale Processing}

Given the extremely low SB nature of Malin 1 disk/spiral arms, after
the basic 
image processing described before (calibration and image combining),
we perform a more sophisticated procedure in order to enhance  
the signal. For this task we applied a 
noise reduction filter based on the {\em total generalized variation
  regularizator}  
\citep{bredies2010}, TGV hereafter. This is similar to an
anisotropic diffusion filter, promoting piecewise smooth images
instead of piecewise constant images as is common with these filters. 
This denoising filter was applied to both $r$ and $g$
images. These were then integrated into a 
false color image, assigning  
$r$ data to the red channel, $g$ data to the blue channel, and
using a weighted average to create a synthetic green channel. 
Later, the images were de-linearized, by applying a histogram
transform which uses the  
rational interpolation mid-tones transfer function described by
\citet{schlick1994}. This provides more contrast  
to low signal features than a standard gamma transform.
To enhance and isolate the structures of large diffuse objects,
pinpoint sources (i.e. stars)  were removed from the images using an
inpainting algorithm based on TGV. This separation between features
is needed, because stars are very  
contrasted objects, and their signal dominates in the wavelet
domain. Direct filtering leads to Gibbs or  
ringing artifacts, or could enhance structures that correspond to
regions of different stellar density, instead of  
enhancing the smooth background signal. Once  
those features are isolated, local contrast is enhanced using the 
starlet transform (also known as  {\em \`a  trous Wavelets}) \citep{stark2010}. 
Basically, the image is decomposed into layers with a redundant
decomposition, isolating structures at certain characteristic
sizes. Wavelet coefficients are modified with multiplicative factors
to enhance features at each layer.
Finally, small and large scale objects were combined again to achieve
the images  
presented in right panels of Figure \ref{fig1}. All the processing
steps were performed on a per-channel basis, and no information was
transferred through them. To allow a better visual inspection of the
image, an inverted and monochrome version is also presented (bottom
panels of Figure \ref{fig1}, and Figures \ref{fig2} and \ref{fig3}).

\subsection{Photometric calibrations and surface brightness} 

In order to assess our SB limit, it is necessary to 
photometrically calibrate the images. We have observed Landolt
standards, and used the photcal package within 
IRAF to obtain the photometric solutions. Then, and
knowing the pixel size (0.16 arcsec pix$^{-1}$), we estimate that the 
SB we reach in our flat-corrected and combined images at $4
\sigma$ is $\sim 26.5$ mag 
arcsec$^{-2}$ in the $r$ band and $\sim 27.5$ mag arcsec$^{-2}$ in the
$g$ band, as observed in regions R3 and R6 labeled in
Figure \ref{fig3}. These values translate to $\sim 28$ $B$ mag
arcsec$^{-2}$. This is $\sim 3 - 4$ mag arcsec$^{-2}$ fainter than the
typical SB reached by surveys and other works to date \citep{moore2006}.   

\section{Results and Discussion}

Figure \ref{fig1} shows several panels where Malin 1 is presented.
Left panels represent the stacked and combined images. Right
panels show the images with the aditional TGV treatment. In the top 
panels, the combined RGB images of the field are presented (see \S
2.1). In the bottom panels, the inverted and monochrome version of the
same image is shown. Some morphological features are
clearly seen, especially in the right panels. Some of them observed
for the first time. 

\subsection{Bulge, spiral arms and stellar density}

\begin{figure*}
%\epsscale{.80}
%\includegraphics[scale=1.0, bb= 0 0 10 10]{imageRGB.eps}
%\includegraphics[scale=0.8]{imageRGB.eps}
%\plotone{final_inv2.jpg}
%\plotfiddle{PSFILE}{VSIZE}{ROTANG}{HSCALE}{VSCALE}{HTRANS}{VTRANS}
%\plotfiddle{sample.eps}{2.6in}{-90.}{32.}{32.}{-250}{225}
%\plotone{f1.pdf}
%\plotfiddle{f1.pdf}{2.6in}{-90}{32.}{32.}{}{}
%\includegraphics[angle=-90,origin=c, scale=0.6]{f1.pdf}
%\includegraphics[angle=-90,origin=c, scale=0.8]{f1.pdf}
\includegraphics[angle=-90,origin=c, scale=0.8]{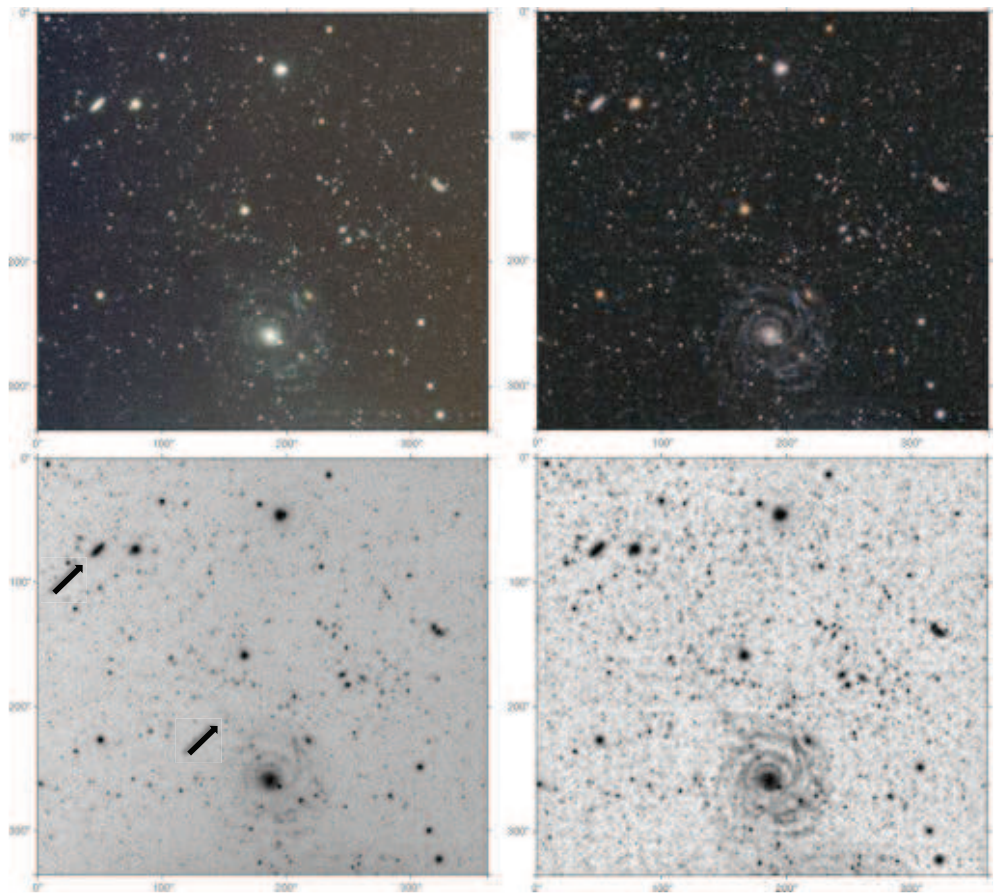}
\caption{Processed images of Malin 1. Left panels: top panel, RGB image
  generated using the stacked and combined $g$ and $r$ images,
  totalizing about 4.5 hrs exposure time. Bottom, the same as the upper panel but a
  monochrome and inverted version of the same image. Right panels: the same as the left
  panels, but after applying 
  the total generalized variation regularizator (TGV), and starlet
  transform for noise reduction and multi-scale 
  processing. The enhancement of the spiral arms and other features discussed in the
  text are apparent after the TGV processing, comparing left and right
  panels. The upper left arrow indicate the galaxy SDSSJ123708.91
  + 142253.2, which is claimed to had interacted 1 Gyr ago with Malin
  1 \citep{reshetnikov2010}. The near center arrow points a
  feature discussed in the text. The data used to create this figure
  is available here. \label{fig1}} 
\end{figure*}

Figure \ref{fig1} shows clear images of a face-on
spiral galaxy presenting a bright inner region which reproduces the
image from \citet{barth2007}. In Figure \ref{fig2} we compare with the
HST/WFPC2 image from \citet{barth2007}. The inner
region of Malin 1 appears much redder 
than any other structure visible in the 
  galaxy. Figures \ref{fig1} and \ref{fig2} show that Malin 1 exhibits
  well formed spiral arms. This demonstrates that in ultra-faint disk
  LSBGs the spiral 
  structure can be well developed to very faint surface brightness
  limits. In order to quantify the density contrast of the
  spiral arms, we can compare the estimated density of these
  structures with the
  stellar density in the solar neighborhood. The MW disk at the solar
  vecinity has a stellar mass surface density of $\sim 49$
  M$_\odot$ pc$^{-2}$ \citep{flynn2006}. From the same author, the
  luminosity surface density is 25.6 L$_\odot$ 
  pc$^{-2}$, or 23.5 mag arcsec$^{-2}$ in the $B$ band. Converting the
  observed SB of $\sim 28$ $B$ mag arcsec$^{-2}$ in luminosity surface
  density in solar units, we obtain $\sim 0.41$ L$_\odot$ pc$^{-2}$, i.e.
  about 62 times lower than the surface luminosity density in 
  the solar neighborhood indicated above\footnote{We have used
    M$_{\odot,B} = 5.45$ mag and a SB for the solar neighborhood of 27.02
    mag arcsec$^{-2}$.}. Using a mass-luminosity
  for our local volume of (M/L)$_B$ = 1.4 \citep{flynn2006}, we obtain
  a stellar mass column density of $\sim 0.57$ M$_\odot$ pc$^{-2}$,
  again, 62 times smaller than the 35.5 
  M$_\odot$ pc$^{-2}$ in the local volume (see also
  \citet{moni2012}). Note that we do not correct these values for the
  effect of inclination on SB since our data suggest that Malin 1 is 
  almost face-on.     

  As shown in all Figures we observe, for the first
  time, in full view, that
  Malin 1 is far from having a featureless low surface brightness 
  disk. It presents rich features
  suggesting stellar formation regions an well defined spiral
  arms. We emphasize that these are well developed structures, with an
  extremely low SB, which translates to extremely diffuse stellar
  density (as shown above). The data then suggests
  that morphological structures as observed here, can be long lived in
  this kind of LSBGs, and could imply large concentrations of dark
  matter \citep{seigar2014}. The formation of spiral arms observed 
  to such low stellar densities is another issue. This could be
  explained by the 
  action of density waves in the matter distribution. However, such a
  theory is still in discussion by some authors
  \citep{choi2015}. Or perhaps a better explanation for the spiral
  structure of Malin 1 is tidal driving or even shear driving
  \citep{dwarkadas1996}, more consistent with the 
  presumed past encounter between Malin 1 and the galaxy
  SDSSJ123708.91 + 142253.2, shown in the images of Figure \ref{fig1}
  by the upper left arrow. 

\subsection{Streams and other features}

 We observe two diffuse and cigar-like features
  embedded in the disk of Malin 1, and seem to cross it
  radially. The shorter one labeled as A in Figure \ref{fig3} is likely the
  disk of a galaxy with a bulge located at RA(J2000.0) 12:36:59 and 
  DEC(J2000) +14:19:44. The SDSS database indicate that this object
  has a photometric redshift of 0.19$\pm0.05$, suggesting that this
  galaxy is in the background, i.e. about 30,000 km/s away from
  Malin 1. Its morphology, undisturbed by the presence of Malin 1
  seems to confirm that its redshift estimate is
  correct. Nevertheless, this value 
  should be confirmed spectroscopically in order to ensure this galaxy
  is not interacting with Malin 1. The other, larger structure,
  labeled as B in Figure \ref{fig3} and also apparent in Figures
  \ref{fig1} and \ref{fig2}, does not present any bulge and its nature  
  is unclear. We do not discard that may be we
  are observing the jet of the active galactic nucleus of Malin
  1, classified as a LINER from spectral data by \citet{barth2007}, and as
  a high-excitation Seyfert galaxy by \citet{impey1989}. The other
  possibility, and perhaps the most likely, is that  
  this feature is part of the claimed past interaction between Malin 1 and
  the galaxy SDSSJ123708.91 + 142253.2 approximately 1 Gyr ago,
  located at $\sim 350$ kpc from Malin 1 \citep{reshetnikov2010}
  (indicated by upper arrows in Figure \ref{fig1}).  We
  note that this 
  stream extends up to $\sim 200$ kpc from the center of
  Malin 1. We do not discard the relationship of these features with
  the presence of Malin 1B (see Figure \ref{fig3}), discussed in
  \citet{reshetnikov2010}.  

\subsection{Stellar formation regions?}

The presence of several diffuse regions in the
  spiral arms exhibiting morphologies similar to stellar formation regions
  and/or clumps of gas is apparent. Examples of these clumps are labeled as R1,
  R2, R3, R4, R5 and R6 in Figure \ref{fig3}. Three of these regions
  (R1, R2 and R3) are the same as those observed by
  \citet{braine2000} with the IRAM 30m telescope for detecting CO,
  unsuccessfully, though. These
  three regions were selected by \citet{braine2000} after
  original observations by \citet{bothun1987}. This is the first time
  that such regions are so well defined in optical images.

\subsection{Optical Diameter of Malin 1 and spiral arms thikness}

\begin{figure*}
%\epsscale{.80}
%\includegraphics[scale=1.0, bb= 0 0 10 10]{test.eps}
%\plotone{features_inv3.jpg}
%\plotone{f2.pdf}
%\includegraphics[angle=-90,origin=c, scale=0.6]{f2.pdf}
%\includegraphics[angle=0,origin=c, scale=0.7]{f2.pdf}
\includegraphics[angle=0,origin=c, scale=0.7]{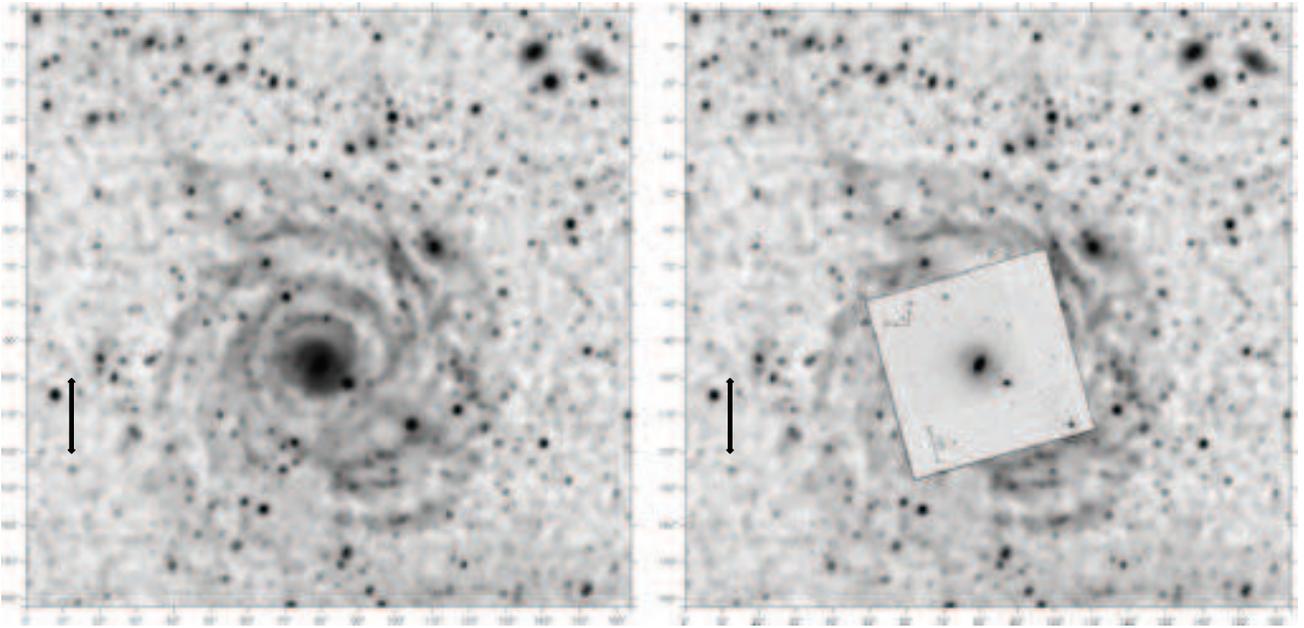}
\caption{Zoom images of the monochrome versions of bottom right panel
  of Figure
  \ref{fig1}. In the right panel, an inset with an HST/WFPC2 image, from \citet{barth2007}, is
  shown. Both panels include the image scale in arcsecs. The double arrows represent the
  physical scale (30.6 kpc, the approximate diameter of the Milky
  Way). \label{fig2}}
\end{figure*}

\begin{figure*}
%\epsscale{.80}
%\includegraphics[scale=1.0, bb= 0 0 10 10]{test.eps}
%\plotone{features_inv3.jpg}
%\plotone{f2.pdf}
%\includegraphics[angle=-90,origin=c, scale=0.6]{f2.pdf}
\includegraphics[angle=0,origin=c, scale=0.7]{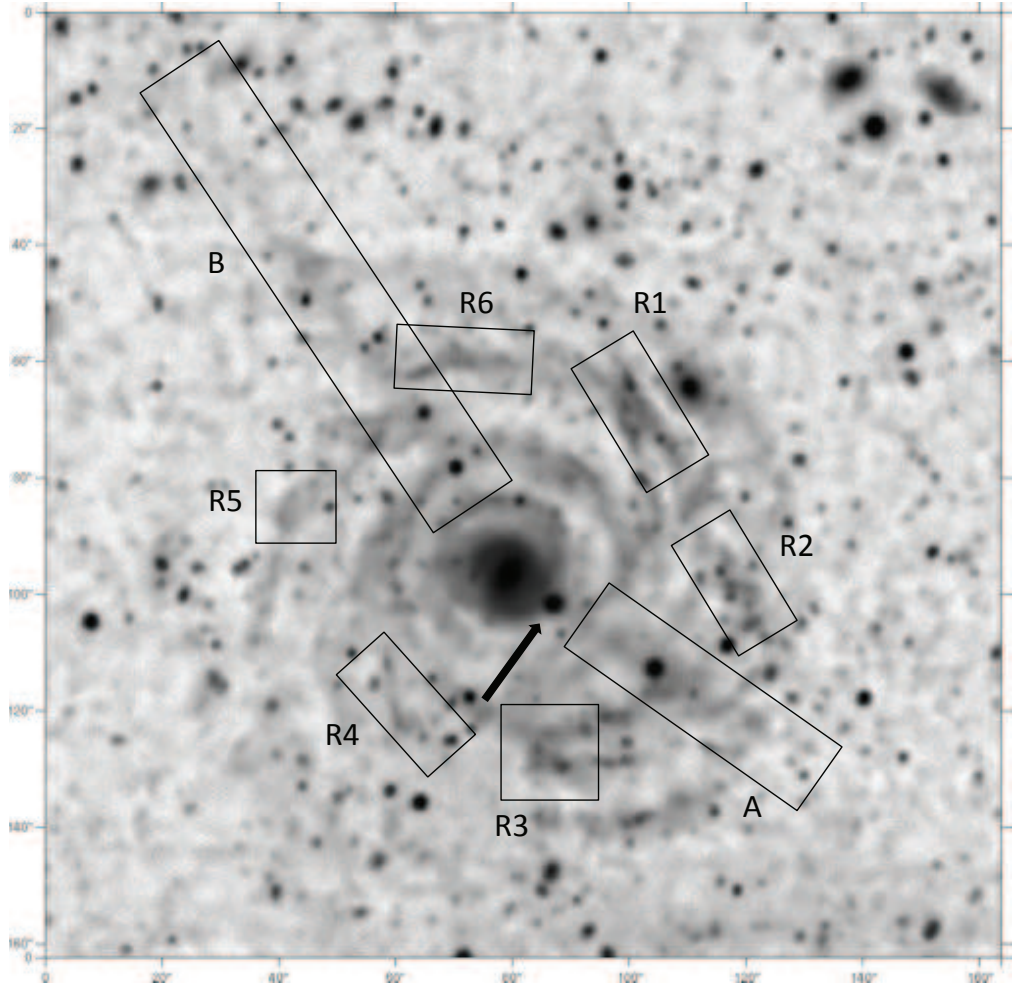}
\caption{The same as left panel of Figure \ref{fig2} but showing some
  regions discussed in the text. The arrow indicate the position of
  Malin 1B \citep{reshetnikov2010}. \label{fig3}}
\end{figure*}

One of the most impressive features of Malin 1 is its large
diameter. \citet{bothun1987, bothun1997} measured an optical diameter
of 110 kpc, almost 6.5 times larger than the diameter of the MW,
making Malin 1 the largest disk/spiral galaxy in the universe so
far. Considering a 
redshift of 0.082, a CCD pixel scale of 0.16 arcsec/pix, and the
faintest spiral arm observed in the images shown in Figures \ref{fig1} and
\ref{fig2}, the optical diameter of Malin 1 turn to be about 
160 kpc (750,000 light years). This is about
50 kpc (163,000 light years) larger than the diameter described by 
former authors \citep{pickering1997, barth2007}. The double arrow in
Figure \ref{fig2} indicate a physical scale of 30.6 kpc, basically the
size of the MW. Images show that Malin 1 is not only large, but also
its spiral arms are scaled to gigantic proportions. Some sections in spiral
arms of Malin 1 in Figure \ref{fig2} measure $\sim 5 - 10$ kpc thick,
i.e. approximately one third of the MW diameter.  

\section{Conclusions}

In this letter we have obtained deep optical images of Malin 1, the
faintest and largest giant LSBG so far, to unprecedented depth. 
The galaxy is composed by an inner bright region, well differentiated
from the outer part composed by spiral arms. The color difference
between the inner part and the spiral arms also appear extreme, with the
former much redder than the spiral arms. 
We clearly see spiral arms and other structures down to a SB $\sim 28$
$B$ mag arcsec$^{-2}$. Comparing the resulting luminosity and stellar
surface density of Malin 1 spiral arms with that derived for the disk of our
Galaxy, we obtain $\sim 0.41$ L$_\odot$ pc$^{-2}$ and $\sim 0.57$
M$_\odot$ pc$^{-2}$, i.e. 62 times lower than those MW corresponding
values. And yet, the galaxy exhibits a textbook and gigantic spiral
structure, to these very low density values. This is certainly the
most astonishing result of this work. How spiral arms can form and
remain long-lived to 
these very low density contrast, is beyond the scope of this
letter. The deep $g$ and $r$ images reveal details for other
features marginally observed in the past. In particular, we observe
two conspicuous features with cigar-like morphology, crossing the spiral
arms: one is very likely a background galaxy, and the other is a
larger, faintest structure pointing to the galaxy SDSSJ123708.91 +
142253.2, claimed by \citet{reshetnikov2010} to had interacted 1 Gyr
ago with Malin 1. Our imaging does not allow to conclude about the nature of
this structure. Other observed features seem to be stellar formation
regions or gas clumps, some of them already observed by
\citet{braine2000} with the IRAM 30m telescope in order to detect CO,
unsuccessfully, however. Our data also confirm that Malin 1 is even
larger than previously reported, reaching 160 kpc diameter, i.e. more
than 6.5 times the diameter of the MW. The fundamental
result is that Malin 1 appears, at last, in full view in optical bands,
allowing to see features that  were not revealed before to such a
detail and contrast.   

\acknowledgments

GG, VS and LI acknowledge the support of Fondecyt regular 1120195 and Basal
PFB-06 CATA. We are greatful to the Las Campanas Observatory staff. We
appreciate discussions with Hugo Messias, Andr\'es Jord\'an, Ren\'e M\'endez and M\'arcio
Catelan. We are greatful to the anonymous referee who
helped to improve this letter.

%We are grateful to the anonymous referee who helped to
%improve this paper. 
.
%\appendix

%\section{Appendix material}




\clearpage

%% the file, uncomment the \plotone, \plottwo, and
%% If you need a layout that cannot be achieved with \plotone or
%% \plottwo, you can invoke the graphicx package directly with the
%% \includegraphics command or use \plotfiddle. For more information,
%% please see the tutorial on "Using Electronic Art with AASTeX" in the
%% documentation section at the AASTeX Web site, http://aastex.aas.org/
%%
%% The examples below also include sample markup for submission of
%% supplemental electronic materials. As always, be sure to check
%% the instructions to authors for the journal you are submitting to
%% for specific submissions guidelines as they vary from

%% This example uses \plotone to include an EPS file scaled to
%% 80% of its natural size with \epsscale. Its caption
%% has been written to indicate that additional figure parts will be
%% available in the electronic journal.

%%% Table

%\begin{table}
%\begin{tabular}{ccccc}
%
%\end{table}

\clearpage

%% Here we use \plottwo to present two versions of the same figure,
%% one in black and white for print the other in RGB color
%% for online presentation. Note that the caption indicates
%% that a color version of the figure will be available online.
%%

%% This figure uses \includegraphics to scale and rotate the still frame

%% If you are not including electonic art with your submission, you may
%% mark up your captions using the \figcaption command. See the
%% No more than seven \figcaption commands are allowed per page,
%% so if you have more than seven captions, insert a \clearpage

%% Tables should be submitted one per page, so put a \clearpage before

%% Two options are available to the author for producing tables:  the
%% deluxetable environment provided by the AASTeX package or the LaTeX
%% table environment.  Use of deluxetable is preferred.
%%

%% Three table samples follow, two marked up in the deluxetable environment,
%% one marked up as a LaTeX table.

%% In this first example, note that the \tabletypesize{}
%% command has been used to reduce the font size of the table.
%% We also use the \rotate command to rotate the table to
%% landscape orientation since it is very wide even at the
%% Note also that the \label command needs to be placed

%% This table also includes a table comment indicating that the full
%% version will be available in machine-readable format in the electronic

\clearpage

\end{document}